\documentclass[conference]{IEEEtran}
\IEEEoverridecommandlockouts
\usepackage{cite}
\usepackage{amsmath,amssymb,amsfonts}
\usepackage{algorithmic}
\usepackage{graphicx}
\usepackage{multirow}
\usepackage{makecell}
\usepackage{textcomp}
\usepackage{xcolor}
\usepackage{colortbl}
\usepackage{booktabs}
\usepackage{comment}
\usepackage{bbding}
\usepackage{threeparttable}
\def\BibTeX{{\rm B\kern-.05em{\sc i\kern-.025em b}\kern-.08em
    T\kern-.1667em\lower.7ex\hbox{E}\kern-.125emX}}
\begin{document}

\title{Text-Video Retrieval With Global-Local Contrastive Consistency Learning 
%{\footnotesize \textsuperscript{*}Note: Sub-titles are not captured in Xplore and
%should not be used}
%\thanks{Identify applicable funding agency here. If none, delete this.}
}

\author{\IEEEauthorblockN{1\textsuperscript{st} Xiaolun Jing}
\IEEEauthorblockA{\textit{Ningbo Artificial Intelligence Institute} \\
\textit{Department of Automation}\\
\textit{Shanghai Jiao Tong University}\\
Shanghai, China \\
jingxiaolun@sjtu.edu.cn}
\and
\IEEEauthorblockN{2\textsuperscript{nd} Xinxing Yang}
\IEEEauthorblockA{\textit{Ningbo Artificial Intelligence Institute} \\
\textit{Department of Automation}\\
\textit{Shanghai Jiao Tong University}\\
Shanghai, China \\
yangxinxing@sjtu.edu.cn}
\and
\IEEEauthorblockN{3\textsuperscript{rd} Genke Yang\IEEEauthorrefmark{1}}
\IEEEauthorblockA{\textit{Ningbo Artificial Intelligence Institute} \\
\textit{Department of Automation}\\
\textit{Shanghai Jiao Tong University}\\
Shanghai, China \\
gkyang@sjtu.edu.cn}
}

\maketitle

\begin{abstract}
Text-video retrieval aims to find the most semantically similar videos with given text queries. However, since videos contain more diverse content than texts, the main semantics expressed by each text-video pair is often partially relevant. The primary methods involve the utilization of language-video attention module to align texts and videos. Though effective, this paradigm inevitably introduces prohibitive computational overhead, resulting in inefficient retrieval. In this paper, we propose a simple yet effective method called Global-Local Contrastive Consistency Learning (GLCCL) to achieve texts and videos semantics alignment. Specifically, we design a parameter-free Global-Local Interaction Module (GLIM) to generate semantic-related frame and video features in a text-guided manner. Furthermore, a Contrastive Score Consistency (CSC) loss is developed to promote consistency learning among different scores on positive pairs and suppress consistency learning on negative pairs. Empirical evidence suggests that CSC loss provides the model with robust discriminative power between positives and hard negatives. Extensive experiments on three benchmark datasets, including MSR-VTT, DiDeMo and VATEX, demonstrate the effectiveness and superiority of our approach.
\end{abstract}

\begin{IEEEkeywords}
Global-Local Interaction, Contrastive Consistency Learning, Text-Video Retrieval
\end{IEEEkeywords}

\section{Introduction}
In recent years, portable filming devices and video media platforms have undergone booming development, resulting in urgent demand for searching videos of interests. Text-video retrieval (TVR) \cite{zhu2023deep, zhang2023multi, fang2022multi} is of great practical value and has raised increasing attention from the academia community. The goal of this task is to align the video candidates with text queries to identify the most semantically relevant videos. Current  approaches mainly benefit from large-scale image-text pre-trained model CLIP \cite{radford2021learning} and achieve promising results on mainstream video-text retrieval benchmarks. For example, CLIP4Clip \cite{luo2022clip4clip} firstly transfers the CLIP knowledge to TVR task and brings significant performance gain. X-Pool \cite{gorti2022x} aggregates the frame features into video feature conditioned on text's attention weights. X-CLIP \cite{ma2022x} utilizes cross-grained contrasts to reduce the negative effects of unnecessary information. DRL \cite{wang2022disentangled} presents the Weighted Token-wise Interaction (WTI) to explore pair-wise correlations. 

%% Figure
%%% Motivation
\begin{figure}[!t]
	\centerline{\includegraphics[width=0.42\textwidth]{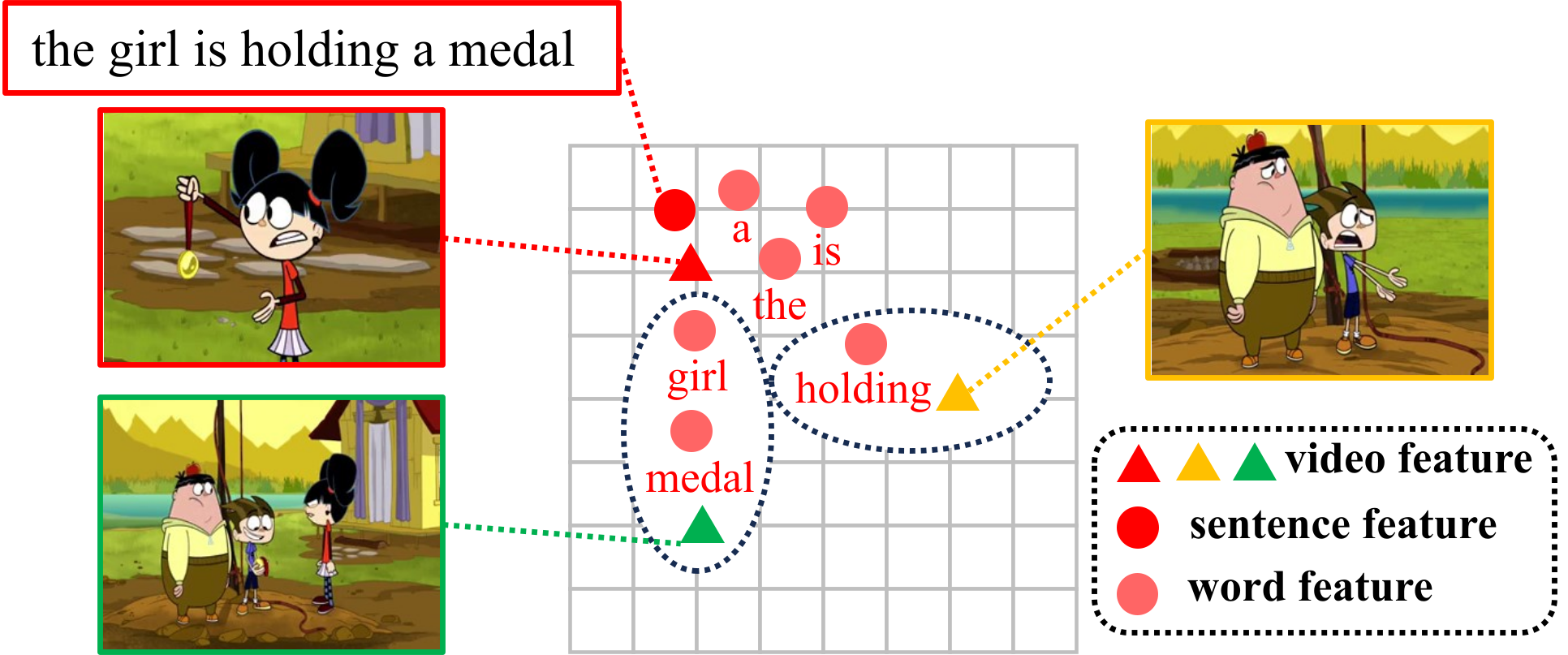}}
	\caption{Illustration of the partially related semantic correspondence between sentence (words) and frames from MSR-VTT. Both textual features only capture sub-regions of frames.}
	\label{fig: motivation}
\end{figure}

In this paper, we study the problem of information asymmetry between texts and videos in text-video retrieval. As illustrated in Fig. \ref{fig: motivation}, the sentence is semantically matched with the top left frame while unrelated to others. As for the fine-grained word features, we surprisingly notice that `holding' is simultaneously presented in the top left and right frames. Similarly, the bottom frame also contains `girl' and `media'. This observation reveals that not limited to the sentence and frame combination, the partial-related phenomenon also exists between words and frames. To address this issue, prior works mainly follow the paradigm of X-Pool \cite{gorti2022x} that uses a cross-modal language-video attention module to generate semantic-relevant video representation. However, this module contains a gamut of trainable parameters with large computational complexity, thus decreasing the inference speed in practical applications. To this end, we propose a Global-Local Interaction Module (GLIM) to generate semantically related frame and video features by the softmax-based weighted combination. Our GLIM has at least two advantages. 1) The computational complexity of GLIM is significantly decreased. 2) GLIM requires no additional trainable parameters. 

Moreover, to facilitate contrastive consistency learning among multi-grained scores, we design an auxiliary  Contrastive Score Consistency (CSC) loss to promote consistency learning on positive pairs and suppress consistency learning on negative pairs. Specifically, we compute the mean variance of multiple scores on positive and negative pairs, respectively, followed by the division to obtain the CSC loss. Due to the great  discriminative power brought by CSC loss, our retrieval model can easily distinguish positive pairs from false positive pairs.

In short, our main contributions are summarized as follows: 

$\bullet$ We propose a parameter-free Global-Local Interaction Module (GLIM) to align text and video semantics with different granularity. 

$\bullet$  We design an auxiliary Contrastive Score Consistency (CSC) loss to promote consistency learning on positive pairs and suppress consistency learning on negative pairs. 

$\bullet$ Our proposed approach achieves comparable results across three public benchmarks of MSR-VTT \cite{xu2016msr}, DiDeMo \cite{anne2017localizing} and VAEX \cite{wang2019vatex}.

\section{Related Works}
\textbf{Vision-Language Pre-Training.} Vision-language understanding is a challenging task that aims to relate and align textual and visual semantics. Recently, with the success of image-text contrastive pre-training \cite{radford2021learning, jia2021scaling} on large-scale web data, this paradigm has shown satisfying results in various downstream tasks, such as VQA \cite{antol2015vqa}, visual reasoning \cite{perez2018film}, image captioning \cite{xu2015show}, \textit{etc}. For video counterparts, video-text pre-training on large-scale video caption datasets, \textit{e.g.} WebVid2M \cite{bain2021frozen} and HowTo100M \cite{miech2019howto100m}, has also significantly boosted video-language understanding.  Nonetheless, video-text pre-training typically requires dense video-text data and huge computation resources. To alleviate this burden, efforts like CLIP4Clip \cite{luo2022clip4clip} and X-CLIP \cite{ma2022x} are presented to transfer the knowledge of image-text pre-training to the video domain, showcasing their remarkable performancce. Thus, our work leverages this scheme for the text-video retrieval task.

\textbf{Text-Video Retrieval.} Text-video retrieval is a fundamental task in the vision-language domain. Early works \cite{liu2019use, gabeur2020multi, dzabraev2021mdmmt} devote to designing dedicated fusion strategies for the cross-modal alignment between offline extracted video and text features. Subsequently, the end-to-end paradigm of receiving raw videos/texts as input has gained large popularity. Frozen \cite{bain2021frozen} presents a dual encoder model for end-to-end training on both image-text and video-text data. ClipBERT \cite{lei2021less} proposes a sparse sampling mechanism on video clips to conduct end-to-end training. TMVM \cite{lin2022text} leverages masked-based prototypes for video features aggregation. Due to the great success of large-scale contrastive image-text pre-training in CLIP \cite{radford2021learning}, some recent works tend to apply CLIP in video-text retrieval. CLIP4Clip \cite{luo2022clip4clip} is the pioneering work to learn better video-text representations from CLIP knowledge. CenterCLIP \cite{zhao2022centerclip} and TS2-Net \cite{liu2022ts2} design a token cluster or token selection module to select the most information tokens. X-Pool \cite{gorti2022x} attempts to aggregate frame features into video representation through a sophisticated language-video attention module. Different from it, we aim to tackle the partially related problem by generating semantic-relevant visual representations in a parameter-free and computationally efficient manner. 

\textbf{Consistency Learning.} The consistency learning has a long history in deep learning. In semi-supervised learning, \cite{laine2016temporal} introduces a self-ensembling method to encourage label consistency between model predictions. CDS \cite{jeong2019consistency} presents a consistency regularization algorithm for object classification and localization. Moreover, consistency learning has been successfully applied to contrastive learning. For example, CGC \cite{pillai2022consistent} devises a contrastive learning method to produce more consistent explanations.
CoCor \cite{wang2023contrastive} employs data augmentation consistency metric to facilitate the systematic integration of diverse data augmentations. Unlike all of them, we mainly explore the multi-grained contrastive consistency learning among multiple scores.

\section{Methodology}
%% Figure
%%% Framework
\begin{figure*}[!t]
	\centerline{\includegraphics[width=0.8\textwidth]{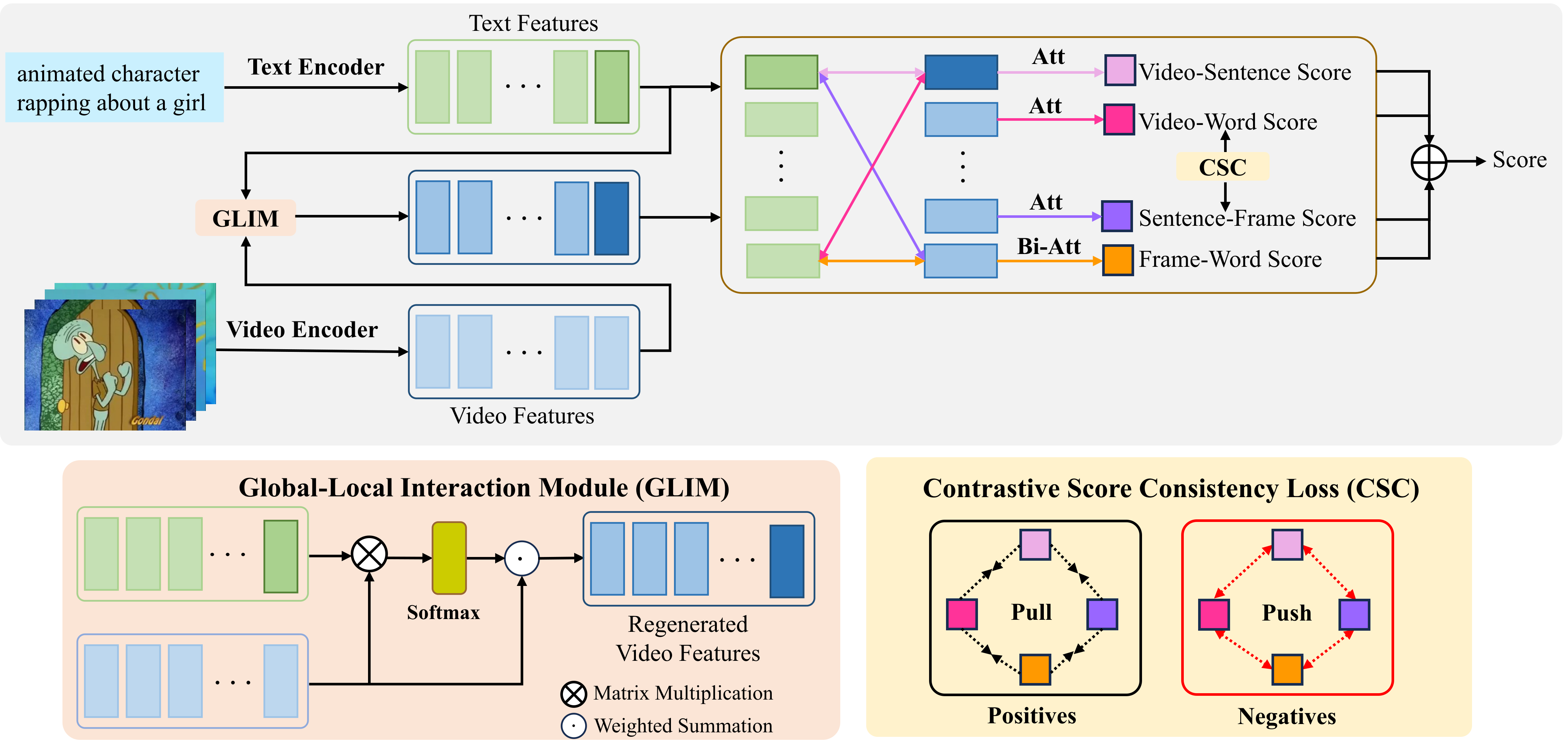}}
	\caption{Our GLCCL framework, containing two key designs: (1) Global-Local Interaction Module (GLIM), which aims to generate text-guided video features with different granularity in a parameter free manner, and (2) Contrastive Score Consistency (CSC), which aims to simultaneously promote positive pairs consistency learning and suppress negative pairs consistency learning. }
	\label{fig: framework}
\end{figure*}

In this section, we present each component of the proposed GLCCL (Fig. \ref{fig: framework}). Starting with an introduction of feature extraction in the Section \ref{sec: feature_extraction}, we then elaborate on the details of global-local interaction module (GLIM) in Section \ref{sec: global-local interaction module}, followed by the multi-grained interaction mechanism in the Section \ref{sec: multi-grained interaction module}. Finally, we describe the contrastive score calculation and total objective function in the Section \ref{sec: contrastive score calculation} and \ref{sec: objectve function}, respectively.

\subsection{Feature Extraction}
\label{sec: feature_extraction}
Given a text-video collection $\mathcal{D} \in \{\mathcal{T}, \mathcal{V}\}$, we first use CLIP as backbone to extract text and video features. For an input text $t=\{t_{i}\}_{i=1}^{N_{t}} \in \mathcal{T}$, we take the outputs of the [EOS] token and corresponding word tokens as sentence feature $t' \in \mathbb{R}^{D}$ and word features $\tilde{t} \in \mathbb{R}^{N_{t} \times D}$. For an input video $v=\{v_{i}\}_{i=1}^{N_{v}} \in \mathcal{V}$, we similarly generate the frame features $\bar{v}' \in \mathbb{R}^{N_{v} \times D}$ from the [CLS] tokens outputs. Here, $N_{t}$, $N_{v}$ and $D$ represent the number of words, the number of frames and the feature dimension. Meanwhile, we adopt a 4-layers of standard transformer blocks to model the temporal relationship in videos, which can be formulated as:
\begin{equation}
	\bar{v}=\text{TransEnc} (\bar{v}' + P),
\end{equation}
where $\bar{v} \in \mathbb{R}^{N_{v} \times D}$ is the final frame features and $P$ is the temporal position embedding.

\subsection{Global-Local Interaction Module}
\label{sec: global-local interaction module}
To enhance text's semantically relevant visual clues with different granularity, we leverage the  global-local interaction module (GLIM) to obtain text-guided video and frame features. As shown in the bottom left part of Fig. \ref{fig: framework}, it computes softmax-based attention weights between sentence (word) and frame features, followed by the aggregation operation to obtain the text-guided video and frame features. The whole process of GLIM can be formulated as:
\begin{align}
	\bar{t}&=\text{Concat} (t', \tilde{t}), \\
	\text{Attention}(\bar{t}, \bar{v})&=\text{Softmax}(\bar{t} \bar{v}^{T}), \\
	\hat{v}&=\text{Attention}(\bar{t}, \bar{v}) \bar{v},
\end{align}
where the beginning token and other tokens of $\hat{v}$ can be regarded as text-guided video feature $v' \in \mathbb{R}^{D}$ and frame features $\tilde{v} \in \mathbb{R}^{N_{t} \times D}$, respectively.

\subsection{Multi-grained Interaction Module}
\label{sec: multi-grained interaction module}
Once the textual features and text-guided visual features are obtained, we employ the multi-grained interaction mechanism to compute instance-level contrastive scores from video-sentence, video-word, sentence-frame and frame-word contrasts.

\textbf{Video-Sentence Contrast.} We directly use matrix multiplication to obtain the  contrastive score $s_{v-s} \in \mathbb{R}^{1}$:
\begin{equation}
	s_{v-s}=(v')^{T} (t').
\end{equation}

\textbf{Video-Word Contrast.} We first use matrix multiplication to compute contrastive vector $\bar{s}_{v-w} \in \mathbb{R}^{1 \times N_{t}}$:
\begin{equation}
	\bar{s}_{v-w}=(v')^{T} (\tilde{t})^{T}.
\end{equation}
Then we adopt the softmax-based weighted combination to obtain the instance-level contrastive score $s_{v-w} \in \mathbb{R}^{1}$:
\begin{equation}
	s_{v-w}=\text{Softmax}(\bar{s}_{v-w}) (\bar{s}_{v-w})^{T}.
\end{equation}

\textbf{Sentence-Frame Contrast.} Similar to video-word contrast, the instance-level contrastive score $s_{s-f} \in \mathbb{R}^{1}$ from sentence-frame contrast can be obtained through:
\begin{align}
	\bar{s}_{s-f}&=(\tilde{v} t')^{T}, \\
	s_{s-f}&=\text{Softmax}(\bar{s}_{s-f}) (\bar{s}_{s-f})^{T}.
\end{align}

\textbf{Frame-Word Contrast.} We use matrix multiplication to compute contrastive matrix $\bar{s}_{f-w} \in \mathbb{R}^{N_{t} \times N_{t}}$, followed by the bi-directional softmax weighted combination to obtain the instance-level contrastive score $s_{f-w} \in \mathbb{R}^{1}$. The whole process is shown as follows:
\begin{align}
	\bar{s}_{f-w}&=(\tilde{v}) (\tilde{t})^{T}, \\
	s_{f-w}&=(\mathcal{A}_{f}(\mathcal{A}_{w}(\bar{s}_{f-w})) + \mathcal{A}_{w}(\mathcal{A}_{f}(\bar{s}_{f-w})))/2,
\end{align}
where $\mathcal{A}_{f}$ and $\mathcal{A}_{w}$ are the frame-level and word-level softmax-based weighted combinations.

\subsection{Contrastive Score Calculation}
\label{sec: contrastive score calculation}
The contrastive score $s(t_{i}, v_{i})$ is represented as the average value of multi-grained contrastive scores:
\begin{equation}
	s(t_{i}, v_{i}) = \text{Mean}(\bar{s}(t_{i}, v_{i})),
\end{equation}
where $\bar{s}(t_{i}, v_{i})=\{s_{v-s}, s_{v-w}, s_{s-f}, s_{f-w}\}$ represents the contrastive scores set.

\subsection{Objective Function}
\label{sec: objectve function}
Given a batch of $B$ text-video pairs, we adopt the symmetric cross-entropy loss to maximize the scores of positive pairs and minimize the scores of negative pairs in a $B \times B$ similarity matrix:
\begin{align}
	\mathcal{L}_{t2v}&=-\frac{1}{B} \sum_{i=1}^{B} \textup{log} \frac{exp(s(t_{i}, v_{i}))}{\sum_{j=1}^{B} exp(s(t_{i}, v_{j}))}, \\
	\mathcal{L}_{v2t}&=-\frac{1}{B} \sum_{i=1}^{B} \textup{log} \frac{exp(s(t_{i}, v_{i}))}{\sum_{j=1}^{B} exp(s(t_{j}, v_{i}))}, \\
	\mathcal{L}_{\text{InfoNCE}}&=\mathcal{L}_{t2v}+\mathcal{L}_{v2t}.
\end{align}

Moreover, we propose a contrastive score consistency (CSC) loss to promote consistency learning on positive pairs and suppress consistency learning on negative pairs. Specifically, we compute the mean variance of multiple scores on positive and negative pairs, respectively, followed by the 
division to obtain $\mathcal{L}_{\textup{CSC}}$:
\begin{align}
	\text{Var}_{pos}&=\frac{1}{B} \sum_{i=1}^{B} \text{Var}(\bar{s}(t_{i}, v_{i})), \\
	\text{Var}_{neg}&=\frac{1}{B^2-B} \sum_{i=1}^{B} \sum_{j=1, j \neq i}^{B} \text{Var}(\bar{s}(t_{i}, v_{j})),\\
	\mathcal{L}_{\textup{CSC}}&=\frac{\text{Var}_{pos}}{\text{Var}_{neg}},
\end{align}
where $\text{Var}(\cdot)$ denotes the variance computation.

The total training loss $\mathcal{L}$ is given by:
\begin{equation}
	\mathcal{L}=\mathcal{L}_{\text{InfoNCE}}+\mathcal{\eta}\mathcal{L}_{\textup{CSC}}, \label{Eq: csc}
\end{equation}
where $\eta$ controls the trade-off between two terms.

\section{Experiment}
\subsection{Experimental Settings}
\textbf{Datasets.} We evaluate the effectiveness of GLCCL on three common text-video retrieval benchmarks. (1) \textbf{MSR-VTT} \cite{xu2016msr} contains 10,000 videos with 20 captions per video. Following \cite{chen2020fine}, we adopt widely-used `Training-9K' split, where 9,000 videos are used for training and 1,000 videos are used for testing. (2) \textbf{DiDeMo} \cite{anne2017localizing} contains 10,000 videos with 40,000 captions. Following previous works \cite{luo2022clip4clip, ma2022x}, all captions of one video are concatenated into a single query during video-paragraph retrieval. (3) \textbf{VATEX} \cite{wang2019vatex} collects 34,991 videos with multiple captions for each. We follow HGR's \cite{chen2020fine} data split. There are 25,991 videos for training, 1,500 videos for validating, and 1,500 videos for testing. 

\textbf{Evaluation Metrics.} For a fair comparison, we use standard video-text retrieval metrics, including Recall at K (R@K, K=1,5,10, higher is better), Median Rank (MdR, lower is better) and Mean Rank (MnR, lower is better). R@K calculates the percentage of correct videos/texts among the top K retrieved videos/texts. MdR and MnR calculate the median and mean rank of groundtruth in the ranking list, respectively. We also take the sum of all R@K as SumR to show the overall retrieval performance.

\textbf{Implementation Details.} Our base model is X-CLIP \cite{ma2022x}. We conduct experiments on 4 NVIDIA GeForce RTX 3090 GPUs using PyTorch. Following \cite{luo2022clip4clip, ma2022x,wang2022disentangled}, we initialize our text and video encoders with the parameters of CLIP (ViT-B/32). We adopt the Adam optimizer \cite{kingma2014adam} with a cosine learning rate schedule \cite{loshchilov2016sgdr}. We set the initial learning rate as 1e-7 for CLIP encoders and 1e-4 for others. The feature dimension is set as 512. Depending on the dataset complexity, we train on MSR-VTT, DiDeMo, and VATEX for 5, 20 and 5 epochs, respectively. The batch size is 128 for all datasets except DiDeMo (64). We configure the word length and frame length as 32, 12 in MSR-VTT and VATEX while 64, 64 in DiDeMo. During training, we set the CSC loss weight $\eta=0.1$ (in Eq. \ref{Eq: csc}). Note that all videos are compressed to 3FPS (Frame Per Second) with width 224 or height 224.

\subsection{Performance Comparison}
% Table
%% Benchmark Results
\begin{table}
	\caption{Text-to-video comparison on MSR-VTT, DiDeMo and VATEX. `$\dag$' denotes re-training. `-' denotes unavailable results. }
	\label{table: Benchmark Results}
	\centering
	\normalsize
	\resizebox{0.5\textwidth}{!}{
		\renewcommand{\arraystretch}{1.02} % 设置行间距为正常值
		\begin{tabular}{c|l|ccccc}
			\Xhline{1px}
			\textbf{Dataset} & \textbf{Method} & \textbf{R@1$\uparrow$} & \textbf{R@5$\uparrow$}  & \textbf{R@10$\uparrow$} & \textbf{MdR$\downarrow$} & \textbf{MnR$\downarrow$} \\
			\hline
			\multirow{10}*{MSR-VTT} & CE \cite{liu2019use} & 20.9 & 48.8 & 62.4 & 6.0 & 28.2 \\
			~ & MMT \cite{gabeur2020multi} & 26.6 & 57.1 & 69.6 & 4.0 & 24.0 \\
			~ & Frozen \cite{bain2021frozen} & 31.0 & 59.5 & 70.5 & 3.0 & - \\
			~ & TMVM  \cite{lin2022text} & 36.2 & 64.2 & 75.7 & 3.0 & -  \\
			~ & MDMMT \cite{dzabraev2021mdmmt} & 38.9 & 69.0 & 79.7 & \textbf{2.0} & 16.5 \\
			~ & $\textup{CLIP4Clip}^{\dag}$ \cite{luo2022clip4clip} & 44.5 & 71.4 & 81.6 & \textbf{2.0} & 15.3 \\
			~ & CenterCLIP \cite{zhao2022centerclip} & 44.2 & 71.6 & 82.1 & \textbf{2.0} & 15.1 \\
			~ & X-Pool \cite{gorti2022x} & 46.9 & 72.8 & 82.2 & \textbf{2.0} & 14.3 \\
			~ & $\textup{X-CLIP}$ \cite{ma2022x} (Base) & 46.1 & 73.0 & 83.1 & \textbf{2.0} & 13.2 \\
			~ & \textbf{GLCCL(Ours)} & \textbf{47.6} & \textbf{73.2} & \textbf{84.0} & \textbf{2.0} & \textbf{13.0} \\
			\hline
			\multirow{8}*{DiDeMo} & CE \cite{liu2019use} & 16.1 & 41.1 & - & 8.3 & 43.7 \\ 
			~ & ClipBERT \cite{lei2021less} & 20.4 & 48.0 & 60.8 & 6.0 & - \\
			~ & Frozen \cite{bain2021frozen} & 34.6 & 65.0 & 74.7 & 3.0 & - \\
			~ & TMVM \cite{lin2022text} & 36.5 & 64.9 & 75.4 & 3.0 & - \\
			~ & $\textup{CLIP4Clip}^{\dag}$ \cite{luo2022clip4clip} & 40.6 & 66.3 & 76.1 & \textbf{2.0} & 19.8 \\
			~ & $\textup{TS2-Net}^{\dag}$ \cite{liu2022ts2} & 42.5 & 69.3 & 77.8 & \textbf{2.0} & 18.8 \\
			~ & $\textup{X-CLIP}^{\dag}$ \cite{ma2022x} (Base) & 44.7 & 72.7 & 80.6 & \textbf{2.0} & 15.9 \\
			~ & \textbf{GLCCL(Ours)} & \textbf{44.9} & \textbf{73.0} & \textbf{82.2} & \textbf{2.0} & \textbf{13.6} \\
			\hline
			\multirow{7}*{VATEX} & HGR \cite{chen2020fine} & 35.1 & 73.5 & 83.5 & 2.0 & - \\
			~ & SUPPORT \cite{patrick2020support} & 44.6 & 81.8 & 89.5 & \textbf{1.0} & - \\
			~ & QB-Norm \cite{bogolin2022cross} & 58.8 & 88.3 & 93.8 & \textbf{1.0} & - \\
			~ & $\textup{CLIP4Clip}^{\dag}$ \cite{luo2022clip4clip} & 58.7 & 89.3 & 94.7 & \textbf{1.0} & 3.7 \\
			~ & $\textup{TS2-Net}^{\dag}$ \cite{liu2022ts2} & 59.5 & \textbf{89.8} & \textbf{95.0} & \textbf{1.0} & \textbf{3.6} \\
			~ & $\textup{X-CLIP}^{\dag}$ \cite{ma2022x} (Base) & 59.1 & 88.9 & 94.2 & \textbf{1.0} & 3.9 \\
			~ & \textbf{GLCCL(Ours)} & \textbf{60.0} & 89.2 & 94.6 & \textbf{1.0} & 5.4 \\
			\Xhline{1px}
		\end{tabular}
	}
\end{table}

We compare GLCCL with recent works on three popular benchmarks and report text-to-video retrieval results in Tab. \ref{table: Benchmark Results}. It can be seen that GLCCL outperforms existing methods on most of the evaluation metrics. On MSR-VTT, GLCCL achieves 47.6 R@1 and 13.0 MnR, surpassing the baseline by +1.5\% and +0.2\% absolute improvements. In addition, compared with recent methods, \textit{i.e.,} CenterClip and X-Pool, our method yields +3.4\% and +0.7\% improvements on R@1, respectively. These results verify the global-local interaction is useful for video-sentence retrieval. We also improve the R@1 metric from 59.1 to 60.0 on VATEX. Similarly, we find that GLCCL even obtains +2.3\% MnR gain on DiDeMo, which shows the importance of global-local interaction in handling longer captions for video-paragraph retrieval.

\subsection{Ablation Study} 
We conduct ablation experiments on MSR-VTT to evaluate the effectiveness of each component in our GLCCL.
 
%% Interaction Ablation
\begin{table}
	\caption{Ablation study of different interaction levels.}
	\label{table: Interaction Ablation}
	\centering
	\huge
	\resizebox{0.5\textwidth}{!}{
		\renewcommand{\arraystretch}{1} % 设置行间距为正常值
		\begin{tabular}{l|ccc|ccc|c}
			\Xhline{2px}
			\multirow{2}*{\textbf{Interaction}} & \multicolumn{3}{c|}{\textbf{Text $\Longrightarrow$ Video}} &  \multicolumn{3}{c|}{\textbf{Video $\Longrightarrow$ Text}} & \multirow{2}*{\textbf{SumR$\uparrow$}}\\
			\cline{2-4} \cline{5-7}
			& R@1$\uparrow$  & R@5$\uparrow$  & R@10$\uparrow$ & R@1$\uparrow$  & R@5$\uparrow$  & R@10$\uparrow$ \\
			\Xhline{1px}
			Global-only & 47.3 & 72.3 & 83.3 & 45.7 & \textbf{74.1} & 81.9 & 404.6 \\
			Local-only & 46.7 & 72.4 & 83.5 & 47.1 & 73.0 & \textbf{82.3} & 405.0 \\
			Global + Local& \textbf{47.6} & \textbf{73.2} & \textbf{84.0} & \textbf{47.2} & 73.0 & \textbf{82.3} & \textbf{407.3} \\
			\Xhline{2px}
		\end{tabular}
	}
\end{table}

\textbf{Effectiveness of GLIM.}  As shown in Tab. \ref{table: Interaction Ablation}, we compare global+local interaction with different variants (\textit{e.g.} global-only and local-only interactions). The first line is the global-only interaction which only uses softmax-based weighted combination to obtain video feature and the second line represents the local-only interaction which regenerates text-guided frame features via word's attention weights. It can be seen that global+local interaction outperforms global-only and local-only interactions. We deem the reason is the global-only and local-only interactions are complementary, thus bringing significant improvements.

%% CSC Loss Ablation
\begin{table}
	\caption{Ablation study of $\mathcal{L}_{\textup{CSC}}$.}
	\label{table: CSC Loss Ablation}
	\centering
	\huge
	\resizebox{0.5\textwidth}{!}{
		\renewcommand{\arraystretch}{1} % 设置行间距为正常值
		\begin{tabular}{l|ccc|ccc|c}
			\Xhline{2px}
			\multirow{2}*{\textbf{Method}} & \multicolumn{3}{c|}{\textbf{Text $\Longrightarrow$ Video}} & \multicolumn{3}{c|}{\textbf{Video $\Longrightarrow$ Text}} & \multirow{2}*{\textbf{SumR$\uparrow$}} \\
			\cline{2-4} \cline{5-7}
			& R@1$\uparrow$  & R@5$\uparrow$  & R@10$\uparrow$ & R@1$\uparrow$  & R@5$\uparrow$  & R@10$\uparrow$\\
			\Xhline{1px}
			w/o $\mathcal{L}_{\textup{CSC}}$ & 46.4 & \textbf{73.4} & 83.6 & \textbf{47.4} & 72.7 & \textbf{82.6} & 406.1 \\
			w/ $\mathcal{L}_{\textup{CSC}}$ & \textbf{47.6} & 73.2 & \textbf{84.0} & 47.2 & \textbf{73.0} & 82.3 & \textbf{407.3} \\
			\Xhline{2px}
		\end{tabular}
	}
\end{table}

\begin{figure}[!t]
	\centerline{\includegraphics[width=0.5\textwidth]{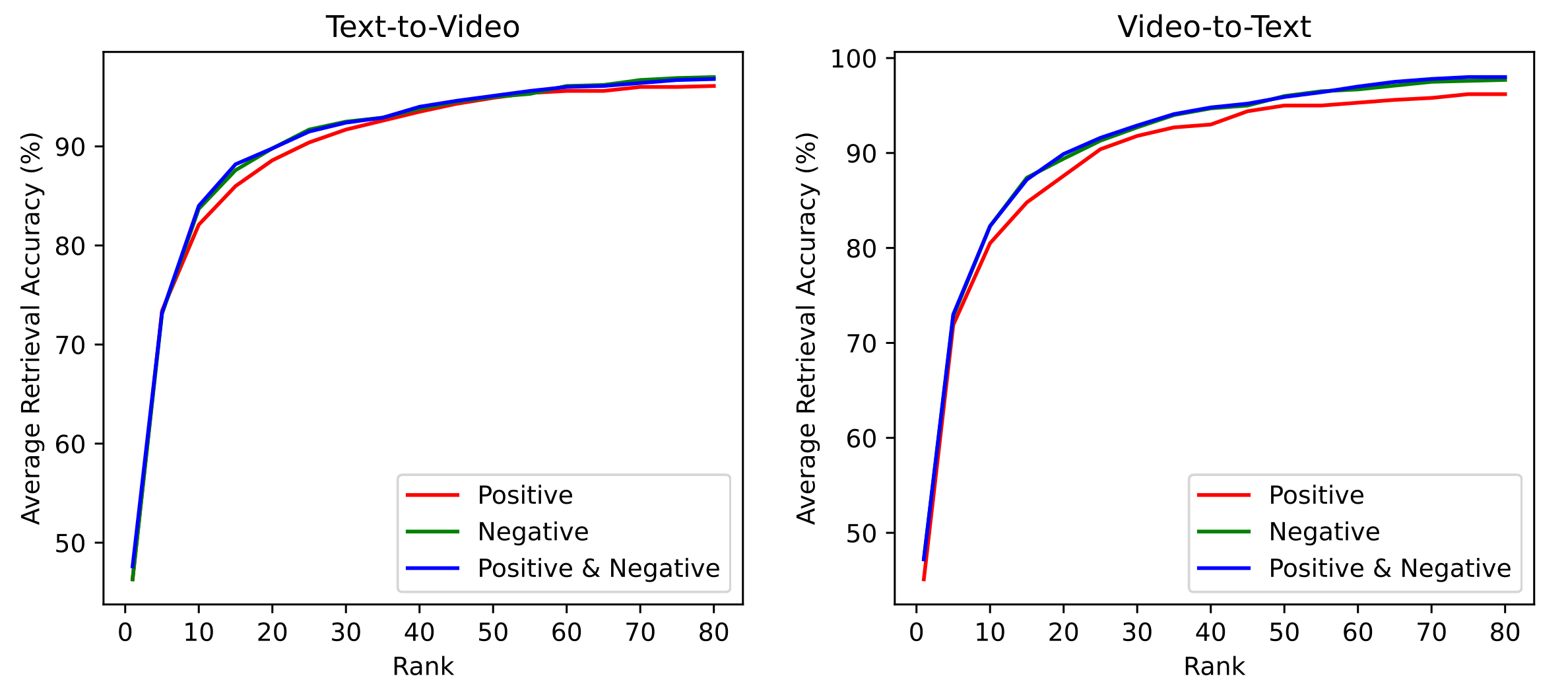}}
	\caption{CMC curves comparison among different variance data in CSC loss. }
	\label{fig: csc_loss_area_ablation}
\end{figure}

\textbf{Effectiveness of CSC Loss. } we first report the results of GLCCL with and without the CSC loss in Tab. \ref{table: CSC Loss Ablation}. It can be observed that the SumR drops 1.2\% without the CSC loss. We think the reason is that CSC loss can significantly enhance the discrimination power between positive and false positive pairs. We also evaluate the impact of different variance data in CSC loss by Cumulative Match Characteristic (CMC) curves. As shown in Fig. \ref{fig: csc_loss_area_ablation}, `Positive' means that the variance of positive pairs represents CSC loss. `Negative' means that CSC loss is given by the variance reciprocal of negative pairs and `Positive \& Negative' denotes our proposed CSC loss. We find that the combination of positives and negatives surpasses each of them, which proves the promotion of positive pairs consistency learning and suppression of negative pairs consistency learning equally contribute to text-video retrieval. 

%% Hyperparameter ablation
\begin{figure}[!t]
	\centerline{\includegraphics[width=0.35\textwidth]{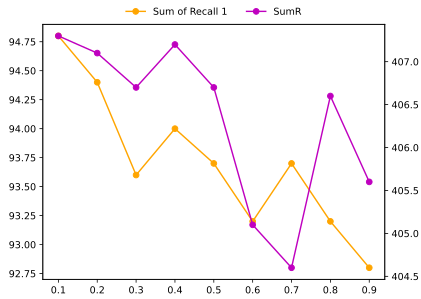}}
	\caption{Effect of hyper-parameter $\eta$ in Eq. \ref{Eq: csc}. }
	\label{fig: csc_loss_weight_ablation}
\end{figure}

\textbf{Impact of hyper-parameter $\eta$. } The parameter $\eta$ aims to trade off $\mathcal{L}_{\textup{InfoNCE}}$ and $\mathcal{L}_{\textup{CSC}}$. We evaluate the scale range setting $\eta \in [0.1, 0.9]$ as shown in Fig. \ref{fig: csc_loss_weight_ablation}. From Fig. \ref{fig: csc_loss_weight_ablation}, the model can get the best Sum of Recall 1 and SumR at $\eta=0.1$. We suppose larger CSC weights may make model deviate from the main optimization direction, thus resulting in retrieval accuracy decline. As a result, we set $\eta=0.1$ as default in practice to achieve the best performance.

%% Computation Efficiency Analysis
\begin{table}[t]
	\centering
	\caption{Computation efficiency comparison on inference time, parameters, FLOPs, and memory usage. Here, the inference time is for the per video evaluation. }
	\label{tab: computation_efficiency}
	\huge
	\resizebox{0.5\textwidth}{!}{
		\begin{tabular}{l*{5}{c}}
			\Xhline{2px}
			\textbf{Method} & \textbf{\makecell[c]{Inference\\Time(ms)$\downarrow$}} & \textbf{Parameters(M)$\downarrow$} & \textbf{FLOPs(G)$\downarrow$} & \textbf{\makecell[c]{Memory\\Usage(M)$\downarrow$}} & \textbf{R@1(\textit{t2v})$\uparrow$} \\
			\hline
			CLIP4Clip\cite{luo2022clip4clip} & 199.6 & 92.62 & \textbf{36.27} & 2869.89 & 44.5 \\
			X-Pool\cite{gorti2022x} & 346.0 & \textbf{85.54} & 37.34 & \textbf{2837.16} & 46.9 \\
			X-CLIP\cite{ma2022x} & \textbf{75.7} & 92.62 & \textbf{36.27} & 2942.48 & 46.1 \\
			\textbf{GLCCL(Ours)} & 182.2 & 92.62 & \textbf{36.27} & 2942.41 & \textbf{47.6} \\
			\Xhline{2px}
		\end{tabular}
	}
\end{table}

\textbf{Computation Efficiency Analysis.} In Tab. \ref{tab: computation_efficiency}, we compare GLCCL with recent CLIP-based methods in terms of efficiency. Note that all models employ CLIP-ViT-B/32 with 64 mini-batch sizes for a fair comparison. Experiments are performed with a 24GB NVIDIA GeForce RTX 3090 GPU. For a more rigorous comparison, we add a column to the Tab. \ref{tab: computation_efficiency} about the comparison of text-to-video R@1. Experimentally, the table reveals two key observations: (i) Compared with X-CLIP, GLCCL achieves better text-to-video R@1 (47.6 \textit{v.s.} 46.1) while introducing no additional parameters and FLOPs. (ii) In terms of the inference time, our GLCCL is slower than X-CLIP but still outperforms other methods, securing the second position. Based on the above analysis, we conclude that GLCCL is an effective and efficient multi-grained alignment framework in text-video retrieval.

\subsection{Qualitative Analysis}
\begin{figure}[!t]
	\centerline{\includegraphics[width=0.41\textwidth]{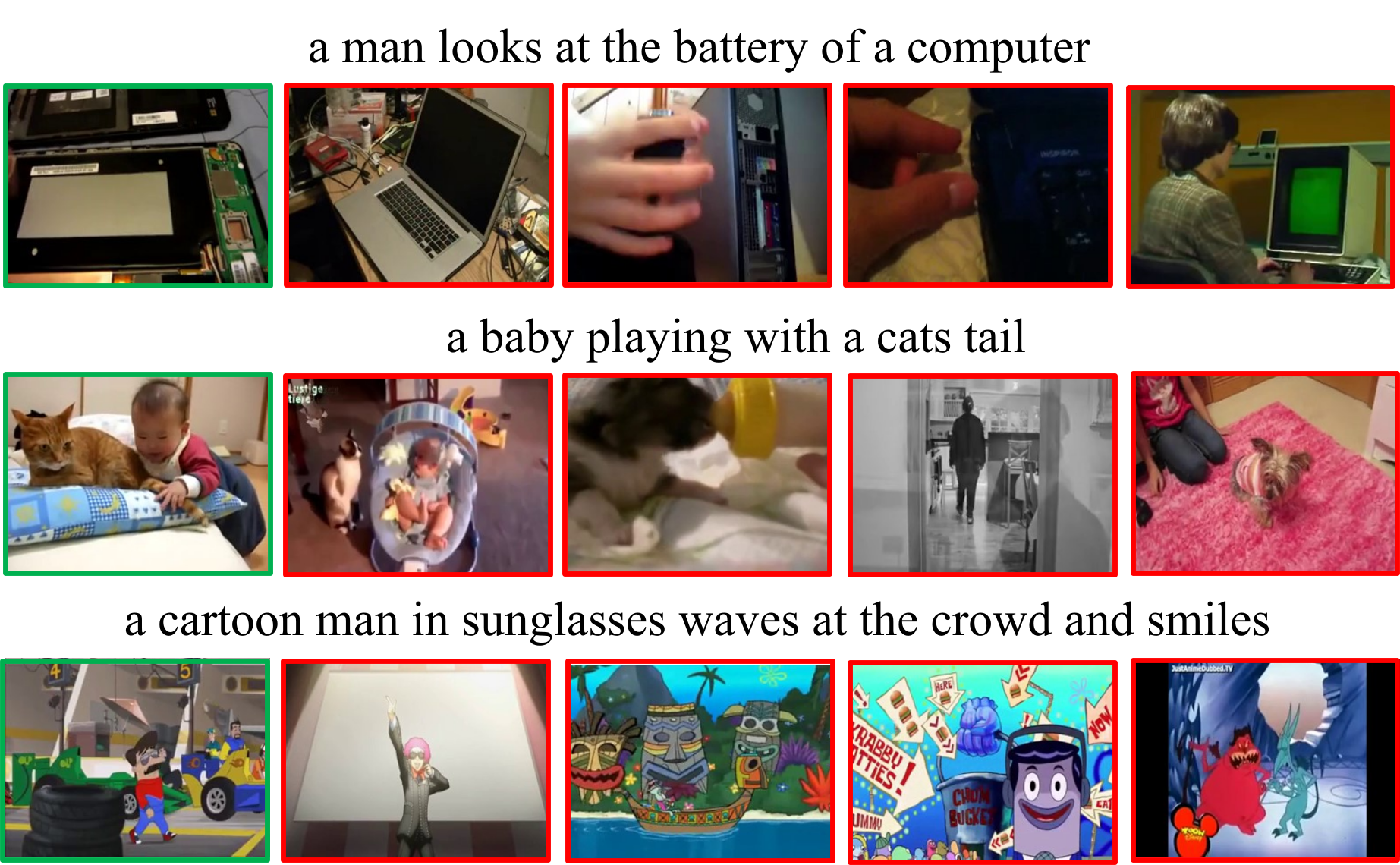}}
	\caption{Visualization of text-to-video retrieval results on MSR-VTT: the top-5 retrieved videos are displayed. Green: ground truth. }
	\label{fig: t2v_visualization}
\end{figure}

\begin{figure}[!t]
	\centerline{\includegraphics[width=0.41\textwidth]{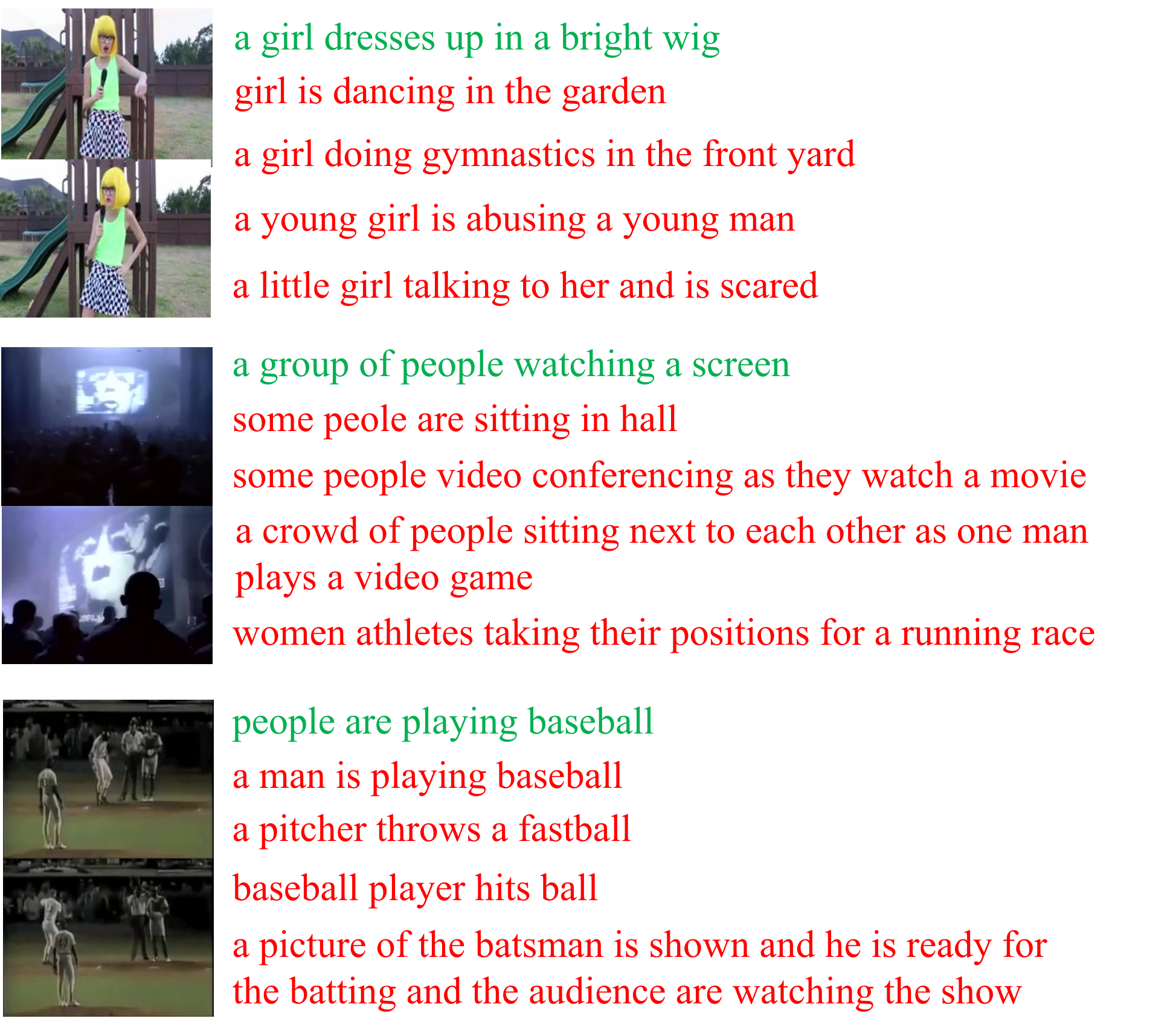}}
	\caption{Visualization of video-to-text retrieval results on MSR-VTT: the top-5 retrieved texts are displayed. Green: ground truth. }
	\label{fig: v2t_visualization}
\end{figure}

To qualitatively validate the effectiveness of our proposed approach, we show 
the text-to-video (T2V) and video-to-text (V2T) retrieval results from MSR-VTT in Fig. \ref{fig: t2v_visualization} and Fig. \ref{fig: v2t_visualization}, respectively. It can be seen that our method successfully retrieves the ground truth based on the given text/video queries. For T2V retrieval in Fig. \ref{fig: t2v_visualization}, in the 1st result, we find that the top-5 results are all about the computer but only the top-1 fully fits the word semantics `battery'. In the 2nd result, only the 1st video is semantic-aligned with `playing with a cats detail' while the others contain partially similar or completely irrelevant semantics. In the 3rd result, we find that all top-5 results contain `cartoon man' but only the 1st video simultaneously captures word semantics `sunglasses' and `crowd'. For V2T retrieval in Fig. \ref{fig: v2t_visualization}, in the 1st and 2nd results, we find that most retrieved results are partially semantic-related but only the top-1 accurately describes the contents of the video, such as semantic details `wig' and `screen'. In the 3rd result, only the 1st text contains all persons in the video while the others contain one person. Both T2V and V2T retrieval results fully demonstrate the merits of global-local interaction module and contrastive scores consistency designs.

%%% t2v visualization

%%% v2t visualization

\section{Conclusion}
This paper presents a novel Global-Local Contrastive Consistency Learning (GLCCL) to address the partially-related problem between texts and videos, and facilitate consistency learning among multi-grained contrastive scores. We attempt to generate text-guided video features with different granularity through a parameter-free Global-Local Interaction Module (GLIM). Besides, we design an auxiliary Contrastive Score Consistency (CSC) loss and incorporate it into the training objective for promoting consistency learning on positive pairs as well as suppressing consistency learning on negative pairs. Quantitative comparisons and qualitative results on three public benchmarks fully demonstrate the robustness, effectiveness and superiority of our proposed GLCCL. In the future, we will extend our method to more sophisticated multi-modal tasks such as video reasoning and video question answering.

\bibliographystyle{IEEEtran}
\bibliography{reference}{}

@article{zhu2023deep,
	title={Deep learning for video-text retrieval: a review},
	author={Zhu, Cunjuan and Jia, Qi and Chen, Wei and Guo, Yanming and Liu, Yu},
	journal={International Journal of Multimedia Information Retrieval},
	volume={12},
	number={1},
	pages={3},
	year={2023},
	publisher={Springer}
}

@inproceedings{zhang2023multi,
	title={Multi-event video-text retrieval},
	author={Zhang, Gengyuan and Ren, Jisen and Gu, Jindong and Tresp, Volker},
	booktitle={ICCV},
	pages={22113--22123},
	year={2023}
}

@article{fang2022multi,
	title={Multi-modal cross-domain alignment network for video moment retrieval},
	author={Fang, Xiang and Liu, Daizong and Zhou, Pan and Hu, Yuchong},
	journal={IEEE Transactions on Multimedia},
	volume={25},
	pages={7517--7532},
	year={2022},
	publisher={IEEE}
}

@inproceedings{xu2016msr,
	title={Msr-vtt: A large video description dataset for bridging video and language},
	author={Xu, Jun and Mei, Tao and Yao, Ting and Rui, Yong},
	booktitle={CVPR},
	pages={5288--5296},
	year={2016}
}

@inproceedings{anne2017localizing,
	title={Localizing moments in video with natural language},
	author={Anne Hendricks, Lisa and Wang, Oliver and Shechtman, Eli and Sivic, Josef and Darrell, Trevor and Russell, Bryan},
	booktitle={ICCV},
	pages={5803--5812},
	year={2017}
}

@inproceedings{wang2019vatex,
	title={Vatex: A large-scale, high-quality multilingual dataset for video-and-language research},
	author={Wang, Xin and Wu, Jiawei and Chen, Junkun and Li, Lei and Wang, Yuan-Fang and Wang, William Yang},
	booktitle={ICCV},
	pages={4581--4591},
	year={2019}
}

@article{liu2019use,
	title={Use what you have: Video retrieval using representations from collaborative experts},
	author={Liu, Yang and Albanie, Samuel and Nagrani, Arsha and Zisserman, Andrew},
	journal={arXiv preprint arXiv:1907.13487},
	year={2019}
}

@inproceedings{gabeur2020multi,
	title={Multi-modal transformer for video retrieval},
	author={Gabeur, Valentin and Sun, Chen and Alahari, Karteek and Schmid, Cordelia},
	booktitle={ECCV},
	pages={214--229},
	year={2020},
	organization={Springer}
}

@inproceedings{dzabraev2021mdmmt,
	title={Mdmmt: Multidomain multimodal transformer for video retrieval},
	author={Dzabraev, Maksim and Kalashnikov, Maksim and Komkov, Stepan and Petiushko, Aleksandr},
	booktitle={CVPR},
	pages={3354--3363},
	year={2021}
}

@inproceedings{antol2015vqa,
	title={Vqa: Visual question answering},
	author={Antol, Stanislaw and Agrawal, Aishwarya and Lu, Jiasen and Mitchell, Margaret and Batra, Dhruv and Zitnick, C Lawrence and Parikh, Devi},
	booktitle={ICCV},
	pages={2425--2433},
	year={2015}
}

@inproceedings{perez2018film,
	title={Film: Visual reasoning with a general conditioning layer},
	author={Perez, Ethan and Strub, Florian and De Vries, Harm and Dumoulin, Vincent and Courville, Aaron},
	booktitle={AAAI},
	volume={32},
	number={1},
	year={2018}
}

@article{xu2015show,
	title={Show, attend and tell: Neural image caption generation with visual attention},
	author={Xu, Kelvin},
	journal={arXiv preprint arXiv:1502.03044},
	year={2015}
}

@inproceedings{miech2019howto100m,
	title={Howto100m: Learning a text-video embedding by watching hundred million narrated video clips},
	author={Miech, Antoine and Zhukov, Dimitri and Alayrac, Jean-Baptiste and Tapaswi, Makarand and Laptev, Ivan and Sivic, Josef},
	booktitle={ICCV},
	pages={2630--2640},
	year={2019}
}

@inproceedings{radford2021learning,
	title={Learning transferable visual models from natural language supervision},
	author={Radford, Alec and Kim, Jong Wook and Hallacy, Chris and Ramesh, Aditya and Goh, Gabriel and Agarwal, Sandhini and Sastry, Girish and Askell, Amanda and Mishkin, Pamela and Clark, Jack and others},
	booktitle={International conference on machine learning},
	pages={8748--8763},
	year={2021},
	organization={PmLR}
}

@inproceedings{jia2021scaling,
	title={Scaling up visual and vision-language representation learning with noisy text supervision},
	author={Jia, Chao and Yang, Yinfei and Xia, Ye and Chen, Yi-Ting and Parekh, Zarana and Pham, Hieu and Le, Quoc and Sung, Yun-Hsuan and Li, Zhen and Duerig, Tom},
	booktitle={ICML},
	pages={4904--4916},
	year={2021},
	organization={PMLR}
}

@inproceedings{bain2021frozen,
	title={Frozen in time: A joint video and image encoder for end-to-end retrieval},
	author={Bain, Max and Nagrani, Arsha and Varol, G{\"u}l and Zisserman, Andrew},
	booktitle={ICCV},
	pages={1728--1738},
	year={2021}
}

@inproceedings{lei2021less,
	title={Less is more: Clipbert for video-and-language learning via sparse sampling},
	author={Lei, Jie and Li, Linjie and Zhou, Luowei and Gan, Zhe and Berg, Tamara L and Bansal, Mohit and Liu, Jingjing},
	booktitle={CVPR},
	pages={7331--7341},
	year={2021}
}

@article{lin2022text,
	title={Text-adaptive multiple visual prototype matching for video-text retrieval},
	author={Lin, Chengzhi and Wu, Ancong and Liang, Junwei and Zhang, Jun and Ge, Wenhang and Zheng, Wei-Shi and Shen, Chunhua},
	journal={NeurIPS},
	volume={35},
	pages={38655--38666},
	year={2022}
}

@inproceedings{chen2020fine,
	title={Fine-grained video-text retrieval with hierarchical graph reasoning},
	author={Chen, Shizhe and Zhao, Yida and Jin, Qin and Wu, Qi},
	booktitle={CVPR},
	pages={10638--10647},
	year={2020}
}

@article{patrick2020support,
	title={Support-set bottlenecks for video-text representation learning},
	author={Patrick, Mandela and Huang, Po-Yao and Asano, Yuki and Metze, Florian and Hauptmann, Alexander and Henriques, Joao and Vedaldi, Andrea},
	journal={arXiv preprint arXiv:2010.02824},
	year={2020}
}

@inproceedings{bogolin2022cross,
	title={Cross modal retrieval with querybank normalisation},
	author={Bogolin, Simion-Vlad and Croitoru, Ioana and Jin, Hailin and Liu, Yang and Albanie, Samuel},
	booktitle={CVPR},
	pages={5194--5205},
	year={2022}
}

@article{luo2022clip4clip,
	title={Clip4clip: An empirical study of clip for end to end video clip retrieval and captioning},
	author={Luo, Huaishao and Ji, Lei and Zhong, Ming and Chen, Yang and Lei, Wen and Duan, Nan and Li, Tianrui},
	journal={Neurocomputing},
	volume={508},
	pages={293--304},
	year={2022},
	publisher={Elsevier}
}

@inproceedings{gorti2022x,
	title={X-pool: Cross-modal language-video attention for text-video retrieval},
	author={Gorti, Satya Krishna and Vouitsis, No{\"e}l and Ma, Junwei and Golestan, Keyvan and Volkovs, Maksims and Garg, Animesh and Yu, Guangwei},
	booktitle={CVPR},
	pages={5006--5015},
	year={2022}
}

@inproceedings{zhao2022centerclip,
	title={Centerclip: Token clustering for efficient text-video retrieval},
	author={Zhao, Shuai and Zhu, Linchao and Wang, Xiaohan and Yang, Yi},
	booktitle={SIGIR},
	pages={970--981},
	year={2022}
}

@inproceedings{liu2022ts2,
	title={Ts2-net: Token shift and selection transformer for text-video retrieval},
	author={Liu, Yuqi and Xiong, Pengfei and Xu, Luhui and Cao, Shengming and Jin, Qin},
	booktitle={ECCV},
	pages={319--335},
	year={2022},
	organization={Springer}
}

@inproceedings{ma2022x,
	title={X-clip: End-to-end multi-grained contrastive learning for video-text retrieval},
	author={Ma, Yiwei and Xu, Guohai and Sun, Xiaoshuai and Yan, Ming and Zhang, Ji and Ji, Rongrong},
	booktitle={Proceedings of the 30th ACM International Conference on Multimedia},
	pages={638--647},
	year={2022}
}

@article{wang2022disentangled,
	title={Disentangled representation learning for text-video retrieval},
	author={Wang, Qiang and Zhang, Yanhao and Zheng, Yun and Pan, Pan and Hua, Xian-Sheng},
	journal={arXiv preprint arXiv:2203.07111},
	year={2022}
}

@article{kingma2014adam,
	title={Adam: A method for stochastic optimization},
	author={Kingma, Diederik P},
	journal={arXiv preprint arXiv:1412.6980},
	year={2014}
}

@article{loshchilov2016sgdr,
	title={Sgdr: Stochastic gradient descent with warm restarts},
	author={Loshchilov, Ilya and Hutter, Frank},
	journal={arXiv preprint arXiv:1608.03983},
	year={2016}
}

@article{laine2016temporal,
	title={Temporal ensembling for semi-supervised learning},
	author={Laine, Samuli and Aila, Timo},
	journal={arXiv preprint arXiv:1610.02242},
	year={2016}
}

@article{jeong2019consistency,
	title={Consistency-based semi-supervised learning for object detection},
	author={Jeong, Jisoo and Lee, Seungeui and Kim, Jeesoo and Kwak, Nojun},
	journal={NeurIPS},
	volume={32},
	year={2019}
}

@inproceedings{pillai2022consistent,
	title={Consistent explanations by contrastive learning},
	author={Pillai, Vipin and Koohpayegani, Soroush Abbasi and Ouligian, Ashley and Fong, Dennis and Pirsiavash, Hamed},
	booktitle={CVPR},
	pages={10213--10222},
	year={2022}
}

@article{wang2023contrastive,
	title={Contrastive learning with consistent representations},
	author={Wang, Zihu and Wang, Yu and Chen, Zhuotong and Hu, Hanbin and Li, Peng},
	journal={arXiv preprint arXiv:2302.01541},
	year={2023}
}

\end{document}